\documentclass[twocolumn,aps]{revtex4}
\usepackage{graphicx}

\usepackage{amssymb}
\usepackage[latin1]{inputenc}
\usepackage{bm}
\newcommand{\tk}{\tilde{k}}
\newcommand{\tw}{\tilde{w}}
\begin{document}

\title{Electron backscattering from dynamical impurities in a Luttinger liquid}
\author{Pablo San-Jos\'e}
 \email{pablo.sanjose@icmm.csic.es}
 \affiliation {Instituto de Ciencia de Materiales de Madrid, CSIC, Cantoblanco, E-28049 Madrid, Spain.}
\author{Francisco Guinea}%
 \email{paco.guinea@icmm.csic.es}
 \affiliation{Instituto de Ciencia de Materiales de Madrid, CSIC, Cantoblanco, E-28049 Madrid, Spain.}
\author{Thierry Martin}%
 \email{martin@cpt.univ-mrs.fr}
 \affiliation{Centre de Physique Th\'eorique, Universit\'e de la M\'editerran\'ee,
Case 907, 13288 Marseille}

\date{\today}

\begin{abstract}
Electron backscattering in a Luttinger liquid with an impurity is investigated in the
presence of zero point motion of the phonon lattice. The impurity can mean either a mass
defect, an elastic defect or a pinning defect. The phonon spectrum is then affected by
the presence of the defect, which enters in the renormalization group equations for
backscattering. The RG equation becomes dependent on the energy cutoff for finite Debye
frequency $\omega_D$ giving rise to finite energy effects. We compute the local density
of states and show how the renormalization group flow is affected by a finite $\omega_D$.
\end{abstract}
\maketitle

\section{Introduction}
It is well known that electron-electron interactions give rise to
Luttinger behavior in one-dimensional metallic systems. Standard
studies assume the existence of an instantaneous interaction
between electrons. Retardation effects need to be included if the
interactions involve excitations, such as phonons, of energies
much lower than the electronic bandwidth. An interesting class of
materials where such excitations are expected to play a role are
the carbon nanotubes\cite{DE00}, where one dimensional features
coexist with significant electron-electron and electron-phonon
interactions.

The effect of the coupling to acoustic phonons on a Luttinger
liquid was studied in\cite{LM94,ML95}. An interesting new feature
is the existence of the Wentzel-Bardeen instability\cite{W51,B51}
for sufficiently large couplings. The analysis of the
backscattering induced by an impurity on the transport properties
of a Luttinger liquid\cite{KF92} has also attracted a great deal
of interest. The extension of this problem to the case of a static
impurity interacting with a coupled electron-phonon system is
discussed in\cite{M95}.

In the present work, we generalize previous work to the case where
the source of backscattering in a one dimensional system has
internal dynamics, and retardation effects have to be taken into
account. We consider, in particular, the effect of a local
modification of the elastic properties of a Luttinger liquid where
the interaction between the electrons and the acoustic phonons is
not negligible. We think that this study can be relevant for the
carbon nanotubes. The electron-phonon coupling in nanotubes has
been estimated to be significant\cite{AG03}, and it is possibly
the origin of superconducting features at low
temperatures\cite{Ketal01,G01}. For sufficiently small radii, the
nanotubes are expected to be close to the Wentzel-Bardeen
instability\cite{ME03}. The repulsive electron-electron
interaction is also large\cite{KBF97,EG98,E99}. Both impurities
and phonons are expected to play an important role in the
transport properties of
nanotubes\cite{Retal00,Betal01,Betal03,Jetal04}. Note also that
the Luttinger liquid characteristics of carbon nanotubes have been
studied by measuring the changes induced by impurities and
contacts on the transport properties\cite{Betal99}.

The next section discusses the model of the bulk system studied in
this paper. The different types of defects considered here are
described in section III. The method of calculation used, based on
the Renormalization Group approach described in\cite{KF92} is
presented in section IV. A discussion and the main conclusions of
the paper can be found in Section V.

\begin{figure}
\includegraphics[width=6cm]{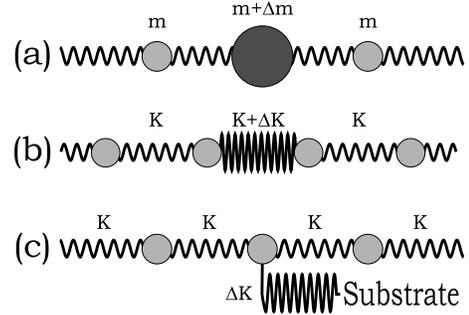}
\caption{a) An impurity atom is inserted in the atomic chain, resulting in a mass defect.
b) a structural deformation of the lattice results in an elastic defect. c) interaction
of one atom with an underlying substrate results in a pinning defect.}
\label{defect}
\end{figure}

\section{The electron-phonon chain.}

Consider a 1-D chain  (length $L$) of $N$ atoms, mass $m$, each pair separated by $a$ and
elastic constant $K$ between each pair. In the absence of inhomogeneities, the action
describing the lattice is written in terms of the Fourier components of the lattice
deformation field $d_{k,\omega}$:
\begin{equation}
S_{ph}=\frac{1}{2}\mu a\sum_{k\omega} \left[ \omega^2-\omega_D^2
\sin^2\left(\frac{k a}{2}\right)\right] d_{k,\omega}d_{-k,-\omega}
\label{S_ph}\end{equation} where $\omega_D$ is the Debye
frequency. The sound velocity is $c=a\omega_D/2=a\sqrt{K/m}$, and $\mu=m/a$ stands for
the linear mass density. We will in the following take the continuum approximation
$4\sin^2(ka/2)/a^2 \rightarrow k^2$.

The charge sector of the electronic degrees of freedom is described as a continuum
Luttinger model (note that phonons and lattice deformations do not couple to the spin
sector). Using a standard bosonized notation, we have:
\begin{equation}
S_{e}=\frac{1}{2}\frac{a}{u K_\rho}\sum_{k\omega} \left(\omega^2-
u^2 k^2 \right) \phi_{k,\omega}\phi_{-k,-\omega}
\label{S_e}\end{equation} Note the analogy between
Eqs.(\ref{S_ph}) and (\ref{S_e}) in the limit of long wavelengths.
One can identify $\mu$ with $(uK_\rho)^{-1}$ and the sound
velocity $c$ with the charge velocity $u$. Usually $u$ is greater
than $c$, since it is of the order of $v_F$ in most cases, but
this need not be the case in the presence of interactions.

We describe the electron-phonon interaction by a deformation
potential:
\begin{eqnarray}
S_{e-ph}&=&-\frac{a}{uK_\rho}b\nu\sum_{k\omega}
k^2\phi_{k,\omega}d_{-k,-\omega} \label{S_e-ph}\end{eqnarray}
where the approximate sign holds for long wavelength excitations.
The reduced coupling strength is defined as $b\equiv
g\sqrt{uK_\rho/\mu}$ and the reduced mass density as $\nu^2 \equiv
\mu u K_\rho$.

The complete model can be then be expressed in terms of a diadic
Green's function:
\begin{eqnarray}
S_{0}&=&\frac{1}{2}\sum_{k\omega}\left[\phi_{-k-\omega},d_{-k-\omega}\right]
{\bf G_{0}}^{-1}(k,\omega)
    \left[\begin{array}{cc}
    \phi_{k\omega} \\
    d_{k\omega}
    \end{array} \right]
\label{diadic_action}
\end{eqnarray}
where the Green's function of this action then reads after inversion:
\begin{eqnarray}
&&{\bf G_{0}}(k,\omega)=\frac{uK_\rho}{2c\omega_D}
\prod_{\epsilon=\pm}
\left(\tilde{\omega}^2-\frac{v_\pm^2}{c^2}\tk^2 \right)^{-1}
\nonumber\\
&&~~~~\times \left[
\begin{array}{cc}
\tilde{\omega}^2-\tk^2 &\frac{bc^2}{\nu} \tk^2\\
\frac{bc^2}{\nu} \tk^2
&\frac{1}{\nu^2}\left(\tilde{\omega}^2-\frac{u^2}{c^2}\tk^2\right)
\end{array}\right]
\label{diadic_green}
\end{eqnarray}
Here $\tk = ( k a ) / 2$ is the dimensionless momentum, $\tilde{\omega}=\omega/\omega_D$
is the dimensionless frequency and $v_\pm\equiv\sqrt{\frac{1}{2}(c^2+u^2\pm\sqrt{4
c^4b^2+(c^2-u^2)^2})}$ are the group velocities of the polarons (hybridized electron and
phonon modes) in the bulk. Note that the Wentzel-Bardeen instability comes about when
$v_-$ becomes imaginary, i.e. for coupling strength $b>u/c$. This completes the
description of the homogeneous electron phonon system.

\section{The defects.}
\subsection{Local change in the mass density.}
We assume that the defect changes locally both the mass density
and the lattice elastic constants. A change in the mass density at
position $x_0$ is described by the action:
\begin{equation}
\Delta S_{m}=\frac{1}{2}\int dt \Delta m (\partial_t
d_{x_0,t})^2=\frac{1}{2}\sum_{\omega}\Delta m \omega^2
d_{x_0\omega}d_{x_0-\omega}
\end{equation}
This results in a modified Green's function which is no longer
translational invariant. In the absence of electron-phonon
coupling, the phonon Green's function becomes:
\begin{eqnarray}
\left.G(x_0,\omega)\right|_{22}&\stackrel{b=0}{=}
\frac{uK_\rho}{2c\omega_D}\frac{1}{\nu^2\tilde{\omega}^2
\sqrt{1-1/\tilde{\omega}^2}+\Delta\nu^2\tilde{\omega}^2}
\end{eqnarray}
with the notation $\Delta\nu^2=u K_\rho \Delta m/a$, which ranges
from $\Delta\nu^2=-\nu^2$ (the mass at position $x_0$ is zero) to
infinity. The retarded, advanced, and causal Green's functions are
obtained by adding a small imaginary part (with the corresponding
sign) to $\omega$, and choosing the principal resolution of the
square root, in which the branch cut is in the negative real axis.

In the absence of a mass defect, we obtain a simple phonon band
for the local density of states. Introducing a finite defect
modifies the phonon spectrum. We can describe the changes induced
by the defect by the transmission coefficient for phonons of
energy $\tw$. The transmission coefficient at low energies is:
\begin{equation}
T ( \tw ) = \frac{1}{1 + i \tw \frac{\Delta m}{m}} \label{T_mass}
\end{equation}
Hence, $\lim_{\tw \rightarrow 0} T ( \tw ) = 1$. The defect is
tranparent for low energy phonons. This result is a consequence of
the fact that the defect does not break the translational
invariance of the lattice. Hence, the system supports long
wavelength phonons of low energy. The deviations from this low
energy limit take place at energies $\tw \sim m / \Delta m$. For
massive defects, $\Delta m \gg m$, the transmission coefficient
tends to zero at finite, but low, energies. The lattice is
effectively divided into two disconnected parts in this range of
energies.

The general case ($b\neq 0$) requires some matrix inversion:
\begin{equation}
G(x_0,\omega)=\left[G_0(x_0,\omega)^{-1}+ \left(\begin{array}{cc}
0&0\\
0&\Delta m \omega^2
\end{array}
\right)\right]^{-1}~. \label{G_mass}
\end{equation}
Analytic expressions can be obtained for all ranges of $\Delta m$. Regardless of the
presence of the defect, the electron band and the phonon bands hybridize, as the phonon
LDOS shows a peak at frequency $u\omega_D/c$, similarly the electron spectrum  shows a
peak at $\omega_D$. The fact that the perturbation enters in eq.(\ref{G_mass}) multiplied
by $\omega^2$ is a sign of its irrelevance at low energies, as discussed earlier.
\subsection{Local change in the elastic constants.}
In the case of an elastic defect,
one modifies the spring constant between two atoms of the chain
$x_0$ and $x_1$.
The action associated with this type of defect then reads
\begin{eqnarray}
\Delta S_{K}&=&-\Delta V=-\frac{1}{2}\int dt \Delta K
(d_{x_1,t}-d_{x_0,t})^2=\\
&=&-\frac{1}{2}\sum_{\omega}\Delta K
[d_{x_0-\omega},d_{x_1-\omega}] \left(
\begin{array}{rr} 1&-1\\-1&1 \end{array}\right)
\left[
\begin{array}{cc} d_{x_0\omega}\\d_{x_1\omega} \end{array}\right]
\nonumber
\label{action_elastic}
\end{eqnarray}

Note that contrary to the case of a mass defect, here two phonon
fields (at locations $x_0$ and $x_1$) enter this action.
In order to obtain the local density of states, we
first derive the $3\times 3$
local Green's function matrix ${\bf G_0}$ associated with
fields $\phi_{x_0,\omega}$,$d_{x_0,\omega}$ and
$d_{x_1,\omega}$ in the absence of the
perturbation:
\begin{eqnarray}
i{\bf G_0}=\left(\begin{array}{ccc}
\langle\phi_{x_0,\omega}\phi_{x_0,-\omega}\rangle &
\langle\phi_{x_0,\omega}d_{x_0,-\omega}\rangle &
\langle\phi_{x_0\omega}d_{x_1-\omega}\rangle \\
\langle d_{x_0,\omega}\phi_{x_0,-\omega}\rangle &
\langle d_{x_0,\omega}d_{x_0,-\omega}\rangle &
\langle d_{x_0,\omega}d_{x_1,-\omega}\rangle \\
\langle d_{x_1,\omega}\phi_{x_0,-\omega}\rangle &
\langle d_{x_1,\omega} d_{x_0,-\omega}\rangle &
\langle d_{x_1,\omega} d_{x_1,-\omega}\rangle
\end{array}\right)
\nonumber\\
\label{3by3green}
\end{eqnarray}
which requires integration of the full Green's function of Eq.
(\ref{diadic_green}) over momentum. Next, Dyson's equation is used
to take the defect into account:
\begin{equation}
{\bf G}={\bf G_0}+{\bf G_0}\cdot {\bf \Delta V} ({\bf 1}-{\bf \Delta V G_0})^{-1}
\cdot {\bf G_0}
\end{equation}
with ${\bf \Delta V}$ the potential matrix associated with the
action (\ref{action_elastic}):
\begin{eqnarray}
{\bf \Delta V}=\Delta K\left(\begin{array}{rrr}
0&0&0\\0&1&-1\\0&-1&1\end{array}\right)~.
\end{eqnarray}
In the absence of electron-phonon coupling,
analytic expressions for the
perturbed phonon Green's function for $d_{x_0\omega}$,
$\left.G\right|_{22}$ are obtained:
\begin{equation}
\left.G\right|_{22}=\frac{uK_\rho}{2c\omega_D}\frac{\left(\frac{K}{\Delta K}+1\right)
-2\left(\tilde{\omega}^2+\tilde{\omega}^2\sqrt{1-1/\tilde{\omega}^2}\right)}
{\left(\frac{K}{\Delta
K}+1\right)\tilde{\omega}^2\sqrt{1-1/\tilde{\omega}^2}+\tilde{\omega}^2}
\end{equation}
The phonon tansmission coefficient at low energies can be written as:
\begin{equation}
T ( \tw ) = \frac{1}{1 - i \frac{\Delta K}{K + \Delta K} \tw^2}
\end{equation}

 When the electron-phonon coupling is finite the
imaginary part of the local Green's function can be negative even
if $\Delta K$ is within its allowed range. This result is a local
counterpart of the Wentzel-Bardeen instability of the extended
version. Similar instabilities have been studied in other strongly
correlated systems\cite{WA84}.

As in the case of a local change in the mass distribution, a local modification of the
elastic constant does not change the value of the transmission coefficient at low energy,
as this perturbation does not alter the translational invariance of the system. The range
of energies for which the transmission coefficient is close to one is $\tw \sim
\sqrt{K/\Delta K}$.
\subsection{Pinning of the lattice.}
We can also consider the case where the lattice is pinned by an
external perturbation. Then, the action due to this defect is:
\begin{eqnarray}
\Delta S_{K} &=&-\Delta V =-\frac{1}{2}\int dt \Delta K
d_{x_0,t}^2 = \nonumber \\ &= &-\frac{1}{2}\sum_{\omega}\Delta K
d_{x_0-\omega} d_{x_0 \omega} \label{action_pinning}
\end{eqnarray}
Using the methods discussed earlier, we find:
\begin{equation}
G(x_0,\omega)=\left[G_0(x_0,\omega)^{-1}+ \left(\begin{array}{cc}
0&0\\
0&\Delta K
\end{array}
\right)\right]^{-1}~. \label{G_pinning}
\end{equation}
In this  case the perturbation is relevant at low energies, and
the transmission coefficient goes to zero at low energies. The
lattice is effectively divided into two decoupled pieces. When the
electron-phonon coupling is zero, the phonon transmission
coefficient at low energies is: \begin{equation} T ( \tw ) =
\frac{\tw}{\tw + i \frac{\Delta K}{2 K}} \end{equation} Above a
crossover energy, $\tw \ge \Delta K / K$, the defect is
transparent to the phonons, and the transmission coefficient
approaches one.
\section{Electron backscattering.}
\subsection{Flow equations.}
The defects considered in the previous section modify the phonon
and electron LDOS. Now we address how each type of defect affects
electron backscattering. We consider here the limit of weak
backscattering only. Then, without loss of generality, the action
corresponding to a short range potential reads:
\begin{eqnarray}
\Delta S_{e}&=&\frac{1}{4}\delta_e\sum_{rs}\int
d\tau\Psi^+_{rsx_0\tau}\Psi_{-rsx_0\tau}=\dots=\\
&=&\delta_e\int d\tau \cos\left(\sqrt{2\pi}\theta_{\rho
x_0\tau}\right)\cos\left(\sqrt{2\pi}\theta_{\sigma
x_0\tau}\right)\nonumber
\end{eqnarray}

Following\cite{M95}, the flow of the electron backscattering
$\delta_e$ induced by the defect obeys the equation:
\begin{eqnarray}
\frac{2}{\delta_e}\frac{d\delta_e}{d l}=1-\Lambda\left(\left.
\mathcal{G}\right|_{1,1}(x_0,\Lambda)+
\left.\mathcal{G}(x_0,-\Lambda)\right|_{1,1}\right)
 \label{flow_eI}
\end{eqnarray}
$\Lambda$ is the cutoff that scales as $\Lambda=e^{-l}\Lambda_0$, and
$\mathcal{G}(x_0,\omega)$ is the Matsubara Green function, which results from performing
a Wick's rotation on the action. After the Wick's rotation, using the a path integral
formalism, the weight of the paths becomes $e^{-\mathcal{S}}$ instead of $e^{iS}$, where
$\mathcal{S}$ is $\mathcal{S}=\sum_{k\omega}[\phi,d]\mathcal{G}^{-1}[\phi,d]^\dag$
instead of $\mathcal{S}=\sum_{k\omega}[\phi,d]G^{-1}[\phi,d]^\dag$ (where $G$ was chosen
in the causal prescription for convergence). With these definitions it can be shown that
$\mathcal{G}(k,\omega)=-G(k,i\omega)$. This quantity is real, due to the fact that
$\mathcal{G}(k,-\omega)=\mathcal{G}(k,\omega)$.
\begin{figure}
\includegraphics[width=8cm]{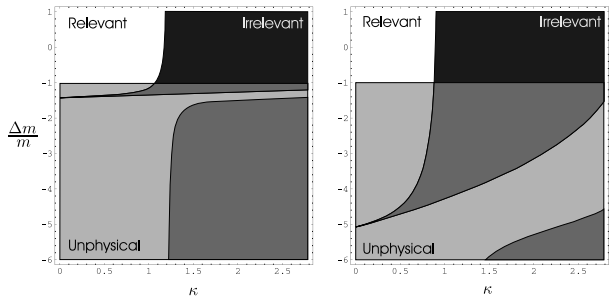}
\caption{Backscattering flow diagram for a mass defect and two
different values of the cutoff, $\Lambda=\omega_D$ (left) and
$\Lambda=0.2\omega_D$. This time the lower grey zone is
unphysical, since it corresponds to a negative substitute mass.
This time the defect resonance always remains in this region.}
\label{RGMass}
\end{figure}

\begin{figure}
\includegraphics[width=7cm]{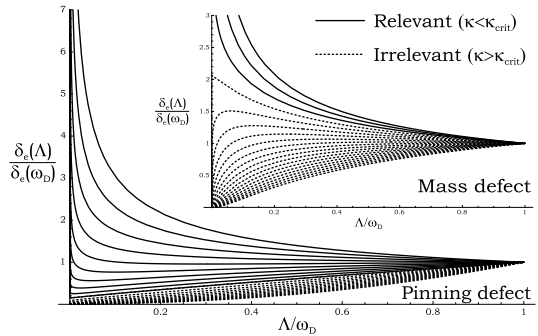}
\caption{Flow diagrams from a given initial backscattering term as function of energy.
Top: defect which changes locally the mass with $\Delta m=1$. Bottom: defect which pins
locally the lattice, with $\Delta K=0.1$. Different curves correspond to different
$\kappa$ electron interaction parameters. Velocities were chosen to be $u=2c$ and $b$ was
set to a constant $2\%$ below the Wentzel-Bardeen instability. }
\label{PinningMassFlow}
\end{figure}

\begin{figure}
\includegraphics[width=8cm]{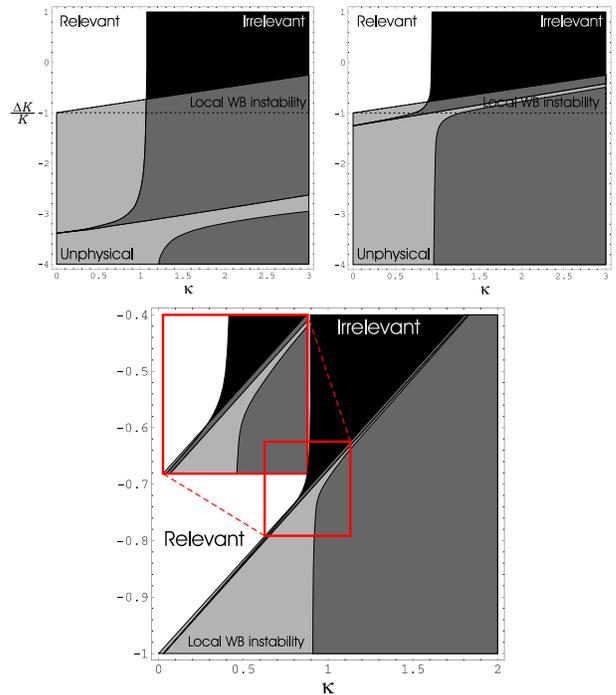}
\caption{Backscattering flow diagram for an elastic defect and three different values of
the cutoff, $\Lambda=\omega_D$ (left), $\Lambda=0.2\omega_D$ (right) and the case of very
small cutoff $\Lambda=0.01\omega_D$ (lower plot). The white and black regions denote
relevant and irrelevant flow of the backscattering respectively, while in the lower grey
zone the system becomes unstable. The local instability boundary (dashed region in the
diagram) corresponds to $\Delta K/K<\Delta
K_\mathrm{crit}/K=-\left(\frac{v_+v_-}{uc}\right)^2$. In the lower plot we have
emphasized the irrelevant backscattering sliver responsible for the effective suppression
at finite energies of elastic defect backscattering close to $\Delta K_\mathrm{crit}$.}
\label{RGElastic1}
\end{figure}

The main difference between the flow of the backscattering term in
eq.(\ref{flow_eI}) and the flow associated with elastic scattering
in Luttinger liquids\cite{KF92} is the non trivial cutoff
dependence in the right hand side in eq.(\ref{flow_eI}).
Retardation effects can also been induced by elastic defects in
systems of electrons coupled to phonons\cite{M95} due to the
difference between the Debye energy, $\omega_D$, and the
electronic bandwidth. If the defect changes the elastic properties
of the lattice, a different crossover at energies lower than
$\omega_D$ is also possible, as discussed in section IIIB.

At the lowest energies or temperatures, a local change in the elastic properties or the
mass density does not affect the phonon transmission coefficient. At these scales, the
defect plays no role, and the scaling of the backscattering term is given by the results
in\cite{M95}. At higher energies or temperatures, although they can be small compared to
$\omega_D$, the phonon transmission coefficient can be significantly reduced, more so the
closer the system is to the bulk Wentzel-Bardeen instability. As the quasiparticles are
made up of a combination of electron and phonon modes, this effect tends to enhance the
electronic backscattering.

\subsection{Mass defects}
The sign of the flow of the backscattering term in eq.(\ref{flow_eI}) allows us to divide
the parameter space into regions where the flow is renormalized towards higher values
(relevant) or towards lower values (irrelevant) at a given energy. These regions are
shown in Fig. [\ref{RGMass}] for a mass defect (see section IIIA). The separatrix between
the two regions tends to be a straight line, independent of $\Delta M$ at low energies,
in agreement with\cite{M95}, where a case equivalent to $\Delta M=0$ was considered. At
higher energies, the region where the backscattering term appears to be relevant is
enlarged. Hence, the flow of the backscattering term is not homogeneous in the regions
where the right hand side of eq. (\ref{flow_eI}) changes sign as function of energy. The
flow of a given initial backscattering term is shown in the upper part of Fig.
[\ref{PinningMassFlow}]. The relevant flow is monotonous, although it deviates
significantly from an exponential dependence on energy \cite{KF92} near the critical
line. The irrelevant region of the paremeter space, which is
$\kappa>\kappa_\mathrm{crit}=\frac{v_+v_-(v_++v_-)}{(c^2+v_+v_-)u}$ is non-monotonous
close to the WB instability, giving an enhanced backscattering of electrons when we are
closed to the $\kappa_\mathrm{crit}$ boundary.

\subsection{Elastic defects}

The flow of the backscattering term is similar when the defect changes the elastic
constants of the lattice (see section IIIB). The main difference is that a local
Wentzel-Bardeen instability can take place even when the modified elastic constants are
positive, and the bulk is stable. The boundary of the locally unstable region is given by
$\frac{\Delta K}{K}<\frac{\Delta
K_\mathrm{crit}}{K}=-\left(\frac{v_+v_-}{uc}\right)^2=-1+ \left(\frac{b}{u/c}\right)^2$,
and is marked by a dashed boundary in Fig. [\ref{RGElastic1}]. Near this $\Delta
K_\mathrm{crit}$ line we find a narrow sliver where the electronic backscattering term
decreases at high energies, although it can eventually be relevant at low energies. This
non monotonous behavior is the inverse of the one discussed for a mass defect. The
asymptotic $\Delta\rightarrow 0$ boundary for relevant-irrelevant behavior of
backscattering on elastic defects is correctly derived in \cite{M95}, and it is exactly
the same as the $\kappa_\mathrm{crit}$ of the mass defect. However at finite temperature
and $\kappa<\kappa_\mathrm{crit}$ (repulsive electrons) backscattering from such defects
can be strongly suppressed when we are close to $\kappa_\mathrm{crit}$ from below and to
the local WB instability at $\Delta K_\mathrm{crit}$. The experimental relevance of this
is greater than the bulk Wentzel-Bardeen instability, since the condition $\Delta
K~\Delta K_\mathrm{crit}$ is much more easily achieved than the bulk instability, and can
have measurable effects on conductance at finite temperatures.
\begin{figure}
\includegraphics[width=7cm]{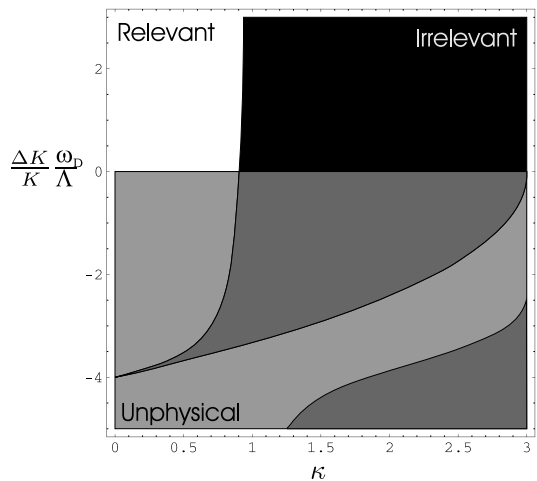}
\caption{Backscattering flow diagram for a defect which elastically pins the lattice at a
given site (due e.g. to the interaction with a substrate).}
\label{FlowPinning}
\end{figure}
\begin{figure}
\includegraphics[width=7cm]{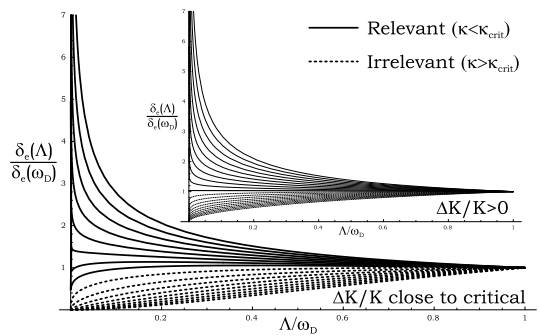}
\caption{Flow diagrams from a given initial backscattering term as function of energy.
The defect changes locally the elastic constant. In the inset $\Delta K>0$ while in the
main plot $\Delta K$ was set to a negative values closer than $1\%$ to the critical
$\Delta K$ which is $\Delta K=-0.4375$ for the parameters used. Velocities are $u=2c$ and
$b=1.5$ is a constant for all curves.}
\label{ElasticFlow}
\end{figure}
Typical flow diagrams of backscattering amplitude for a given initial value as function
of energy and for different bulk parameters are given in Fig. [\ref{ElasticFlow}]. One
can clearly see how the non monotonous behavior discussed above is strongly enhanced near
the critical line (main figure).
\subsection{Pinning of the lattice.}
A representative phase diagram for the case when the lattice is pinned by a defect which
breaks translational invariance is shown in Fig. [\ref{FlowPinning}]. In this case, and
at finite frequencies, the relevant region is reduced, giving rise to a non-monotonous
renormalization of the backscattering amplitude that can be suppressed in a similar way
as a mass defect could be enhanced at finite frequencies. This effect, as in the case of
a substitute mass, is stronger the closer the system gets to the bulk instability. The
flow of a given initial backscattering term is shown in the lower part of
Fig.[\ref{PinningMassFlow}] for a system close to the Wentzel-Bardeen instability. Note
moreover that the pinning defect case presents a peculiar scaling of the flow equations,
which only depend on $\Delta_K/\Lambda$ (see vertical axis of Fig. [\ref{FlowPinning}]).

\section{Conclusions.}
We have analyzed the role of lattice defects in Luttinger liquids
with significant interactions between electrons and acoustical
phonons. The quasiparticles of the system are combinations of
electron-hole pairs and phonons. A lattice defect modifies locally
the propagation of the phonons, inducing a change in the
electronic transport properties as well.

A lattice defect which does not break the translational invariance of the lattice is
irrelevant at low energies, as it is transparent to low frequency phonons. The borderline
which separates the regions where electronic backscattering is relevant or irrelevant is
not affected by a defect of this type, and depends only on bulk parameters\cite{M95}.
There is a range of energies, which can be much lower than the Debye frequency,
$\omega_D$, where the defect scatters the phonons strongly. In this region of energies or
temperatures, the electronic backscattering can be significantly enhanced. Thus the flow
of an electronic backscattering term can be non monotonous, unlike the case of a
Luttinger liquid with elastic scattering\cite{KF92}, giving rise to measurable
conductance dependence on temperature, for example. Even in a system with attractive
interactions, a mass defect can block effectively the transport over a significant range
of temperatures. On the other hand in a system with repulsive electronic interactions an
elastic defect that is close enough to the local Wentzel-Bardeen instability can prove to
be unexpectedly transparent to electrons. This local instability can be induced by a
strong enough local softening of the elastic constant of the chain in the presence of
electron-phonon interactions.
\section{Acknowledgements.}
We are thankul to J. Gonz\'alez for many helpful comments. P. S.
J. and F. G. acknowledge financial support from MCyT (Spain)
through grant MAT2002-0495-C02-01.

\bibliography{Biblio}

\begin{thebibliography}{20}
\expandafter\ifx\csname natexlab\endcsname\relax\def\natexlab#1{#1}\fi
\expandafter\ifx\csname bibnamefont\endcsname\relax
  \def\bibnamefont#1{#1}\fi
\expandafter\ifx\csname bibfnamefont\endcsname\relax
  \def\bibfnamefont#1{#1}\fi
\expandafter\ifx\csname citenamefont\endcsname\relax
  \def\citenamefont#1{#1}\fi
\expandafter\ifx\csname url\endcsname\relax
  \def\url#1{\texttt{#1}}\fi
\expandafter\ifx\csname urlprefix\endcsname\relax\def\urlprefix{URL }\fi
\providecommand{\bibinfo}[2]{#2}
\providecommand{\eprint}[2][]{\url{#2}}

\bibitem[{\citenamefont{Dresselhaus and Eklund}(2000)}]{DE00}
\bibinfo{author}{\bibfnamefont{M.~S.} \bibnamefont{Dresselhaus}}
  \bibnamefont{and} \bibinfo{author}{\bibfnamefont{P.~C.}
  \bibnamefont{Eklund}}, \bibinfo{journal}{Adv. in Phys.}
  \textbf{\bibinfo{volume}{49}} (\bibinfo{year}{2000}).

\bibitem[{\citenamefont{Loss and Martin}(1994)}]{LM94}
\bibinfo{author}{\bibfnamefont{D.}~\bibnamefont{Loss}} \bibnamefont{and}
  \bibinfo{author}{\bibfnamefont{T.}~\bibnamefont{Martin}},
  \bibinfo{journal}{Phys. Rev. B} \textbf{\bibinfo{volume}{50}},
  \bibinfo{pages}{12160} (\bibinfo{year}{1994}).

\bibitem[{\citenamefont{Martin and Loss}(1995)}]{ML95}
\bibinfo{author}{\bibfnamefont{T.}~\bibnamefont{Martin}} \bibnamefont{and}
  \bibinfo{author}{\bibfnamefont{D.}~\bibnamefont{Loss}},
  \bibinfo{journal}{Int. J. Mod. Phys. B} \textbf{\bibinfo{volume}{9}},
  \bibinfo{pages}{495} (\bibinfo{year}{1995}).

\bibitem[{\citenamefont{Wentzel}(1951)}]{W51}
\bibinfo{author}{\bibfnamefont{G.}~\bibnamefont{Wentzel}},
  \bibinfo{journal}{Phys. Rev.} \textbf{\bibinfo{volume}{83}},
  \bibinfo{pages}{168} (\bibinfo{year}{1951}).

\bibitem[{\citenamefont{Bardeen}(1951)}]{B51}
\bibinfo{author}{\bibfnamefont{J.}~\bibnamefont{Bardeen}},
  \bibinfo{journal}{Rev. Mod. Phys.} \textbf{\bibinfo{volume}{23}},
  \bibinfo{pages}{261} (\bibinfo{year}{1951}).

\bibitem[{\citenamefont{Kane and Fisher}(1992)}]{KF92}
\bibinfo{author}{\bibfnamefont{C.~L.} \bibnamefont{Kane}} \bibnamefont{and}
  \bibinfo{author}{\bibfnamefont{M.~P.~A.} \bibnamefont{Fisher}},
  \bibinfo{journal}{Phys. Rev. Lett.} \textbf{\bibinfo{volume}{68}},
  \bibinfo{pages}{1220} (\bibinfo{year}{1992}).

\bibitem[{\citenamefont{Martin}(1995)}]{M95}
\bibinfo{author}{\bibfnamefont{T.}~\bibnamefont{Martin}},
  \bibinfo{journal}{Physica D} \textbf{\bibinfo{volume}{83}},
  \bibinfo{pages}{216} (\bibinfo{year}{1995}).

\bibitem[{\citenamefont{\'Alvarez and Gonz\'alez}(2003)}]{AG03}
\bibinfo{author}{\bibfnamefont{J.~V.} \bibnamefont{\'Alvarez}}
  \bibnamefont{and}
  \bibinfo{author}{\bibfnamefont{J.}~\bibnamefont{Gonz\'alez}},
  \bibinfo{journal}{Phys. Rev. Lett.} \textbf{\bibinfo{volume}{91}}
  (\bibinfo{year}{2003}).

\bibitem[{\citenamefont{Kociak et~al.}(2001)\citenamefont{Kociak, Kasumov,
  Gu\'eron, Reulet, Khodos, Gorbatov, Volkov, Vaccarini, and
  Bouchiat}}]{Ketal01}
\bibinfo{author}{\bibfnamefont{M.}~\bibnamefont{Kociak}},
  \bibinfo{author}{\bibfnamefont{A.~Y.} \bibnamefont{Kasumov}},
  \bibinfo{author}{\bibfnamefont{S.}~\bibnamefont{Gu\'eron}},
  \bibinfo{author}{\bibfnamefont{B.}~\bibnamefont{Reulet}},
  \bibinfo{author}{\bibfnamefont{I.~I.} \bibnamefont{Khodos}},
  \bibinfo{author}{\bibfnamefont{Y.~B.} \bibnamefont{Gorbatov}},
  \bibinfo{author}{\bibfnamefont{V.~T.} \bibnamefont{Volkov}},
  \bibinfo{author}{\bibfnamefont{L.}~\bibnamefont{Vaccarini}},
  \bibnamefont{and} \bibinfo{author}{\bibfnamefont{H.}~\bibnamefont{Bouchiat}},
  \bibinfo{journal}{Phys. Rev. Lett.} \textbf{\bibinfo{volume}{86}},
  \bibinfo{pages}{2416} (\bibinfo{year}{2001}).

\bibitem[{\citenamefont{Gonz\'alez}(2001)}]{G01}
\bibinfo{author}{\bibfnamefont{J.}~\bibnamefont{Gonz\'alez}},
  \bibinfo{journal}{Phys. Rev. Lett.} \textbf{\bibinfo{volume}{87}}
  (\bibinfo{year}{2001}).

\bibitem[{\citenamefont{Martino and Egger}(2003)}]{ME03}
\bibinfo{author}{\bibfnamefont{A.~D.} \bibnamefont{Martino}} \bibnamefont{and}
  \bibinfo{author}{\bibfnamefont{R.}~\bibnamefont{Egger}},
  \bibinfo{journal}{Phys. Rev. B} \textbf{\bibinfo{volume}{67}}
  (\bibinfo{year}{2003}).

\bibitem[{\citenamefont{Kane et~al.}(1997)\citenamefont{Kane, Balents, and
  Fisher}}]{KBF97}
\bibinfo{author}{\bibfnamefont{C.}~\bibnamefont{Kane}},
  \bibinfo{author}{\bibfnamefont{L.}~\bibnamefont{Balents}}, \bibnamefont{and}
  \bibinfo{author}{\bibfnamefont{M.~P.~A.} \bibnamefont{Fisher}},
  \bibinfo{journal}{Phys. Rev. Lett.} \textbf{\bibinfo{volume}{79}},
  \bibinfo{pages}{5086} (\bibinfo{year}{1997}).

\bibitem[{\citenamefont{Egger and Gogolin}(1998)}]{EG98}
\bibinfo{author}{\bibfnamefont{R.}~\bibnamefont{Egger}} \bibnamefont{and}
  \bibinfo{author}{\bibfnamefont{A.}~\bibnamefont{Gogolin}},
  \bibinfo{journal}{Eur. Phys. J. B} \textbf{\bibinfo{volume}{3}},
  \bibinfo{pages}{781} (\bibinfo{year}{1998}).

\bibitem[{\citenamefont{Egger}(1999)}]{E99}
\bibinfo{author}{\bibfnamefont{R.}~\bibnamefont{Egger}},
  \bibinfo{journal}{Phys. Rev. Lett.} \textbf{\bibinfo{volume}{83}},
  \bibinfo{pages}{5547} (\bibinfo{year}{1999}).

\bibitem[{\citenamefont{Reulet et~al.}(2000)\citenamefont{Reulet, Kasumov,
  Kociak, Deblock, Khodos, Gorbatov, Volkov, Journet, and Bouchiat}}]{Retal00}
\bibinfo{author}{\bibfnamefont{B.}~\bibnamefont{Reulet}},
  \bibinfo{author}{\bibfnamefont{A.~Y.} \bibnamefont{Kasumov}},
  \bibinfo{author}{\bibfnamefont{M.}~\bibnamefont{Kociak}},
  \bibinfo{author}{\bibfnamefont{R.}~\bibnamefont{Deblock}},
  \bibinfo{author}{\bibfnamefont{I.~I.} \bibnamefont{Khodos}},
  \bibinfo{author}{\bibfnamefont{Y.~B.} \bibnamefont{Gorbatov}},
  \bibinfo{author}{\bibfnamefont{V.~T.} \bibnamefont{Volkov}},
  \bibinfo{author}{\bibfnamefont{C.}~\bibnamefont{Journet}}, \bibnamefont{and}
  \bibinfo{author}{\bibfnamefont{H.}~\bibnamefont{Bouchiat}},
  \bibinfo{journal}{Phys. Rev. Lett.} \textbf{\bibinfo{volume}{85}},
  \bibinfo{pages}{2829} (\bibinfo{year}{2000}).

\bibitem[{\citenamefont{Bozovic et~al.}(2001)\citenamefont{Bozovic, Bockrath,
  Hafner, Lieber, Park, and Tinkham}}]{Betal01}
\bibinfo{author}{\bibfnamefont{D.}~\bibnamefont{Bozovic}},
  \bibinfo{author}{\bibfnamefont{M.}~\bibnamefont{Bockrath}},
  \bibinfo{author}{\bibfnamefont{J.~H.} \bibnamefont{Hafner}},
  \bibinfo{author}{\bibfnamefont{C.~M.} \bibnamefont{Lieber}},
  \bibinfo{author}{\bibfnamefont{H.}~\bibnamefont{Park}}, \bibnamefont{and}
  \bibinfo{author}{\bibfnamefont{M.}~\bibnamefont{Tinkham}},
  \bibinfo{journal}{Appl. Phys. Lett.} \textbf{\bibinfo{volume}{78}},
  \bibinfo{pages}{3693} (\bibinfo{year}{2001}).

\bibitem[{\citenamefont{Bozovic et~al.}(2003)\citenamefont{Bozovic, Bockrath,
  Hafner, Lieber, Park, and Tinkham}}]{Betal03}
\bibinfo{author}{\bibfnamefont{D.}~\bibnamefont{Bozovic}},
  \bibinfo{author}{\bibfnamefont{M.}~\bibnamefont{Bockrath}},
  \bibinfo{author}{\bibfnamefont{J.~H.} \bibnamefont{Hafner}},
  \bibinfo{author}{\bibfnamefont{C.~M.} \bibnamefont{Lieber}},
  \bibinfo{author}{\bibfnamefont{H.}~\bibnamefont{Park}}, \bibnamefont{and}
  \bibinfo{author}{\bibfnamefont{M.}~\bibnamefont{Tinkham}},
  \bibinfo{journal}{Phys. Rev. B} \textbf{\bibinfo{volume}{67}}
  (\bibinfo{year}{2003}).

\bibitem[{\citenamefont{Javey et~al.}(2004)\citenamefont{Javey, Guo, Paulsson,
  Wang, Mann, Lundstrom, and Dai}}]{Jetal04}
\bibinfo{author}{\bibfnamefont{A.}~\bibnamefont{Javey}},
  \bibinfo{author}{\bibfnamefont{J.}~\bibnamefont{Guo}},
  \bibinfo{author}{\bibfnamefont{M.}~\bibnamefont{Paulsson}},
  \bibinfo{author}{\bibfnamefont{Q.}~\bibnamefont{Wang}},
  \bibinfo{author}{\bibfnamefont{D.}~\bibnamefont{Mann}},
  \bibinfo{author}{\bibfnamefont{M.}~\bibnamefont{Lundstrom}},
  \bibnamefont{and} \bibinfo{author}{\bibfnamefont{H.~J.} \bibnamefont{Dai}},
  \bibinfo{journal}{Phys. Rev. Lett.} \textbf{\bibinfo{volume}{92}}
  (\bibinfo{year}{2004}).

\bibitem[{\citenamefont{Bockrath et~al.}(1999)\citenamefont{Bockrath, Cobden,
  Lu, Rinzler, Smalley, Balents, and McEuen}}]{Betal99}
\bibinfo{author}{\bibfnamefont{M.}~\bibnamefont{Bockrath}},
  \bibinfo{author}{\bibfnamefont{D.~H.} \bibnamefont{Cobden}},
  \bibinfo{author}{\bibfnamefont{J.}~\bibnamefont{Lu}},
  \bibinfo{author}{\bibfnamefont{A.~G.} \bibnamefont{Rinzler}},
  \bibinfo{author}{\bibfnamefont{R.~E.} \bibnamefont{Smalley}},
  \bibinfo{author}{\bibfnamefont{L.}~\bibnamefont{Balents}}, \bibnamefont{and}
  \bibinfo{author}{\bibfnamefont{P.~L.} \bibnamefont{McEuen}},
  \bibinfo{journal}{Nature} \textbf{\bibinfo{volume}{397}},
  \bibinfo{pages}{598} (\bibinfo{year}{1999}).

\bibitem[{\citenamefont{Wu and Anderson}(1984)}]{WA84}
\bibinfo{author}{\bibfnamefont{C.~C.} \bibnamefont{Wu}} \bibnamefont{and}
  \bibinfo{author}{\bibfnamefont{P.~W.} \bibnamefont{Anderson}},
  \bibinfo{journal}{Phys. Rev. B} \textbf{\bibinfo{volume}{29}}
  (\bibinfo{year}{1984}).

\end{thebibliography}
\end{document}